\documentclass[aps,floats,prl,
twocolumn]{revtex4}

\usepackage{amsfonts,amsmath,mathrsfs}
\usepackage{amssymb} \usepackage{dsfont}
\usepackage{bm} \usepackage{bbm}
\usepackage{dcolumn}
\usepackage{epsfig} \usepackage{latexsym}
\usepackage{float}

\begin{document}

\title{Wave thermalization and its implications for nonequilibrium statistical mechanics}

\author{Liyi Zhao$^{1\dagger}$, Ping Fang$^{1\dagger}$ and Chushun Tian$^{2*}$}

\affiliation{$^1$Institute for Advanced Study, Tsinghua University, Beijing 100084, China\\
$^2$CAS Key Laboratory of Theoretical Physics and Institute of Theoretical Physics,
Chinese Academy of Sciences, Beijing 100190, China}

\date{\today}

\begin{abstract}

Understanding the rich spatial and temporal structures in nonequilibrium thermal environments is a major subject of statistical mechanics. Because universal laws, based on an ensemble of systems, are mute on an individual system, exploring nonequilibrium statistical mechanics and the ensuing universality in individual systems has long been of fundamental interest.
Here, by adopting the wave description of microscopic motion, and combining the recently developed eigenchannel
theory and the mathematical tool of the concentration of measure, we show that in a single complex medium, a universal spatial structure -- the diffusive steady state -- emerges from an overwhelming number of scattering eigenstates of the wave equation. Our findings suggest a new principle, dubbed ``the wave thermalization'', namely, a propagating wave undergoing complex scattering processes can simulate nonequilibrium thermal environments, and exhibit macroscopic nonequilibrium phenomena.

\end{abstract}

\maketitle

In standard statistical mechanics, hypothesizing an ensemble of systems (e.g., initial states \cite{Prigogine62} and disorder configurations \cite{Kamenev}) is crucial for formulating physical laws universal with respect to system's details. On the other hand, in reality one often deals with an individual system, where the ensemble does not exist. This raises the fundamental issue of universality and statistical mechanics in individual systems. Although the investigation was initiated in the context of quantum mechanics long time ago \cite{von Neumann29}, only recently, motivated by experiments on individual quantum systems, have substantial progresses been made for equilibrium statistical mechanics \cite{Lebowitz06,Popescu06,Deutsch91,Srednicki94,Rigol08,Rigol16}. It is even more challenging to explore nonequilibrium statistical mechanics and the ensuing universality in individual systems. Indeed, at the core of nonequilibrium statistical mechanics are various spatial and temporal structures \cite{Reichl}, with the diffusive steady state as a prototype. They have not yet been explored for individual systems, owing to the difficulty of treating microscopic motions.

\begin{figure}[h]
\includegraphics[width=8.0cm] {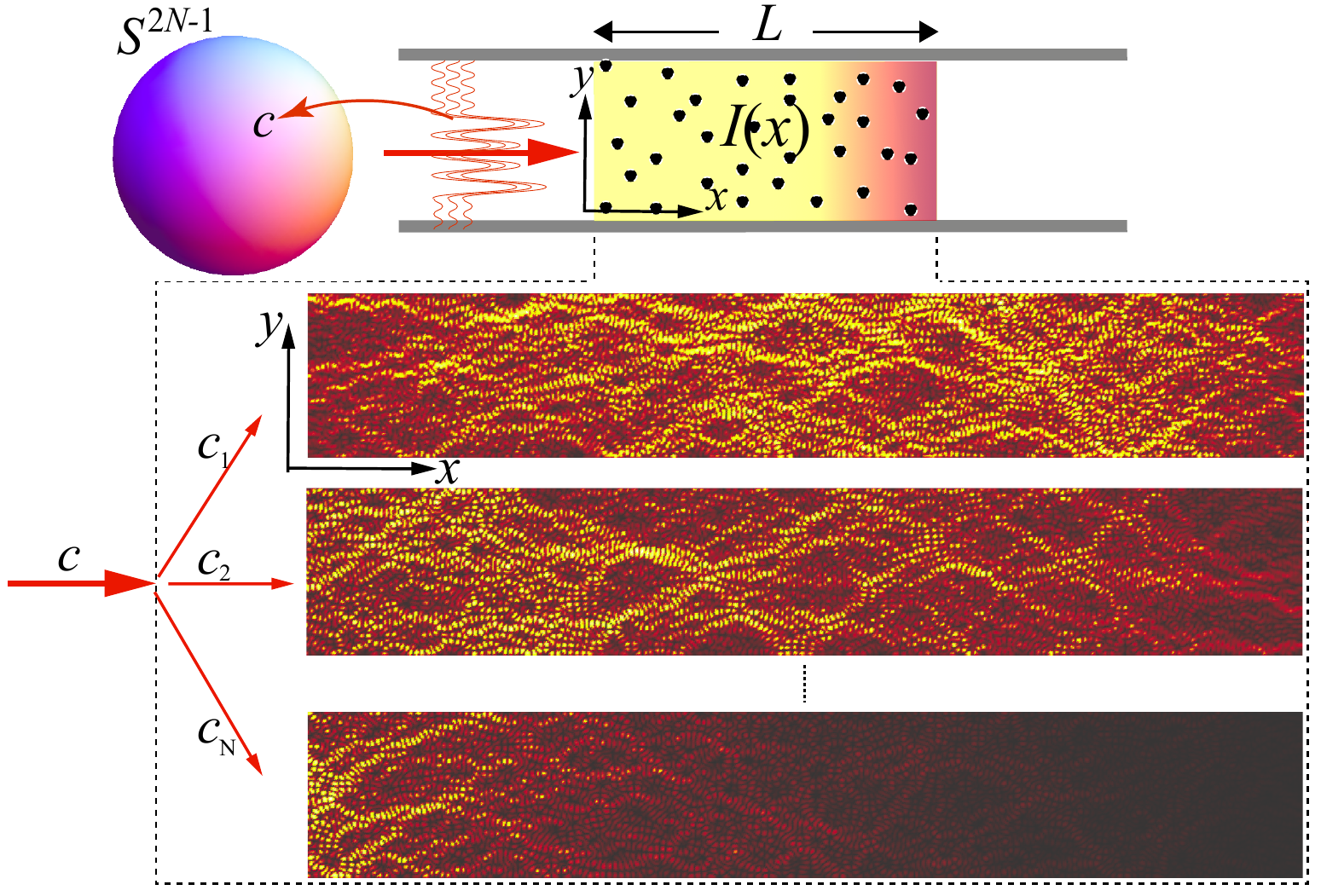}
\caption{In a disordered medium, an incoming wave $c$ of frequency $\Omega$ is projected onto $N$ eigenchannels,
each of which has a specific spatial structure (lower part). For large $N$, this structure is universal with respect to disorder configurations,
and an overwhelming number of incoming waves
behave the same: they pump energies into distinct channels with equal weight, and excite a diffusive steady state $I(x)$ universal with respect to $c$ and disorder configurations (upper part).}
\label{fig:4}
\end{figure}

Here we show that in an individual, complex wave system, an overwhelming number of scattering eigenstates can simulate nonequilibrium thermal environments, even though the propagation of waves is deterministic. As a result, a universal large-scale spatial structure -- a diffusive steady state -- arises, with the wave energy falling linearly in space. This diffusive phenomenon is of wave origin, in contrast to the canonical mechanism of diffusion, i.e., Brownian motion of particles. Moreover, it is attributed to general properties of waves, but not to their specific type, and thus occurs to a variety of waves. In particular, for the quantum matter wave, it is not its probability aspect -- which other waves do not have -- that is responsible for this emergent diffusion.

\begin{figure*}
\includegraphics[width=16.0cm] {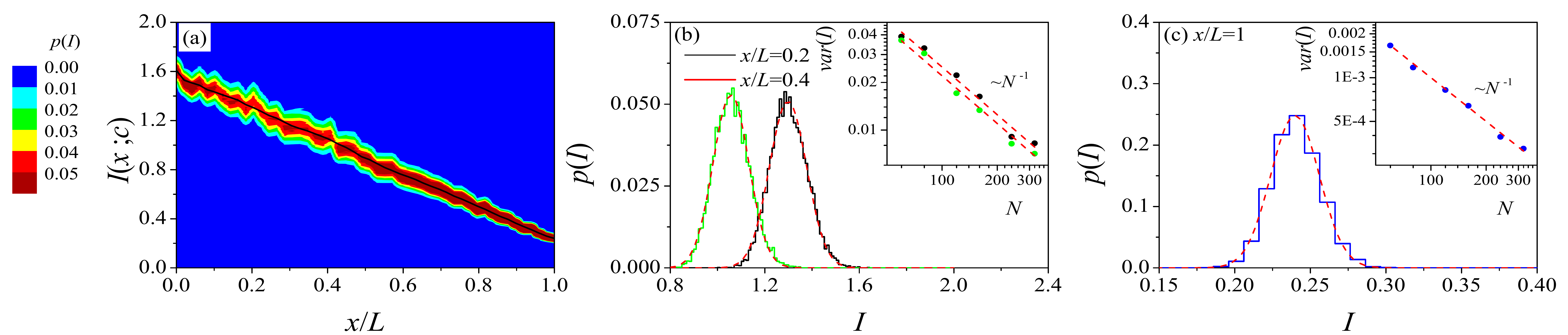}
\caption{(a) We simulate the propagation of $10^4$ randomly chosen incoming waves in a single disordered medium,
with $L=50$ in unit of the inverse wave number (in the ideal waveguide)
and $N=400$. The profiles of
$I(x;c)$ concentrate around $I(x)$ given by Eq.~(\ref{eq:12}) and falling linearly (solid line). The distribution $p(I(x;c))$ is represented by distinct colors. (b,c) Simulated $p(I(x;c))$ (histogram) is well fit by a Gaussian distribution (dashed line) at given $x$ (main panel), with variance $\sim N^{-1}$ (inset).}
\label{fig:1}
\end{figure*}

The theory developed differs fundamentally from standard nonequilibrium statistical mechanics \cite{Prigogine62,Dorfman} in several aspects. (i) It is based on the wave description of microscopic motion, rather than the commonly adopted (quasi)particle description, which is the short wavelength limit of the former. (ii) Besides being more general, the wave description has a great advantage over the particle description: by reducing the microscopic motion into wave propagation in certain channels, it enables a good control of the complexity of microscopic motion, physically and technically. (iii) The study of nonequilibrium phenomena in turn amounts to the analysis of the dependence of observables on the projection of a wave onto these channels and of channels' structures: this is very different from dealing with the ensemble distribution or the density matrix in canonical theories. By undertaking the tasks of (iii), we connect nonequilibrium statistical mechanics to the concentration of measure phenomenon in mathematics \cite{Milman87}, which has found remarkable applications in quantum physics recently \cite{Popescu06,Guarneri15}.

We first discuss qualitatively the results and the implied new principle, {\it the wave thermalization} (cf.~Fig.~\ref{fig:4}).
We focus on the classical scalar wave \cite{Sheng95} below. All discussions and results can be generalized to other waves. Consider a monochromatic wave of (circular) frequency $\Omega$ launched from an ideal waveguide into a disordered dielectric configuration (upper part). The background of disordered medium and the ideal waveguide have the same refractive index of unity. In the disordered medium there are a set of ``natural'' channels, the (transmission) eigenchannels \cite{Choi11,Tian15,Rotter17} (lower part). Each of them has a specific spatial structure characterizing how energies are distributed in it, and a transmission eigenvalue characterizing its capability of transmitting waves (to be discussed in details later). The projection of the incoming wave onto these channels is represented by $N$ (the channel number $\sim$ the area of cross section) complex coefficients: $(c_1,c_2,\cdots,c_N)\equiv c$, with $\sum_{n=1}^N |c_n|^2=1$. They constitute the coordinate of the unit hypersphere $S^{2N-1}$. Physically, this projection describes the decomposition of incoming wave into a number of ``partial waves'' (lower part), which propagate in distinct eigenchannels and superpose to give the wave field. With this projection, an observable is a real function $O(c): S^{2N-1}\rightarrow \mathds{R}$. For large $N$, $S^{2N-1}$ exhibits an intriguing phenomenon of high-dimensional geometry -- the concentration of measure \cite{Milman87}: its entire surface area is almost concentrated around the equator. This results in a concentration property on $O(c)$. Namely, {\it $O(c)$ with nice continuity properties strongly concentrates around some constant value.} Therefore, for an overwhelming number of $c$, $O(c)$ is essentially its $c$-averaging. Alternatively, the averaging can be realized by making the medium in contact with reservoir, that prepares an ensemble of incoming waves. As such, a scattered wave can simulate a system coupled to nonequilibrium thermal environments.

As shown below, a desired $O(c)$ is the energy density. Applying the concentration of measure to it, we find that an overwhelming number of incoming waves behave the same: they pump energies into distinct eigenchannels with equal weight, and a diffusive steady state $I(x)$ universal with respect to $c$, which falls linearly in the longitudinal $x$-direction, results (Fig.~\ref{fig:4}). Moreover, we find that for large $N$, even in a single disorder configuration, eigenchannels have universal spatial and spectral structures, leading to the independence of $I(x)$ on disorder configurations. So, unlike in traditional mesocopic transport theories \cite{Kamenev,Sheng95,Tian15,Dorokhov84,Mello88}, neither a reservoir nor an ensemble of disorder configurations is required here.

Now we turn to the microscopic theory. The wave propagation is described by the Helmholtz equation \cite{Sheng95},
\begin{eqnarray}\label{eq:15}
    \left\{\nabla^2 +\Omega^2 [1+\chi_{[0,L]}(x)\delta\epsilon (x,y)]\right\}E(x,y)=0.
\end{eqnarray}
Here $E(x,y)$ is the wave field. $\chi_{[0,L]}(x)$ takes the value $1$ in the interval $[0,L]$ and $0$ otherwise. $\delta\epsilon (x,y)$ is the disorder part of the dielectric configuration. The group velocity in the background is unity. For a scattering eigenstate, $E(x,y)$ satisfies the boundary condition:
\begin{eqnarray}
    E(x,y)=\Big\{\begin{array}{c}
                    {\rm incoming\, wave}\, + {\rm reflected\, wave},\quad x<0; \\
                    {\rm transmitted\, wave},\quad x>L,\quad\quad\quad\quad\,
                  \end{array}
                  \nonumber
\end{eqnarray}
where the incoming wave is normalized such that it corresponds to a unit energy flux.

\begin{figure*}
\includegraphics[width=16.0cm] {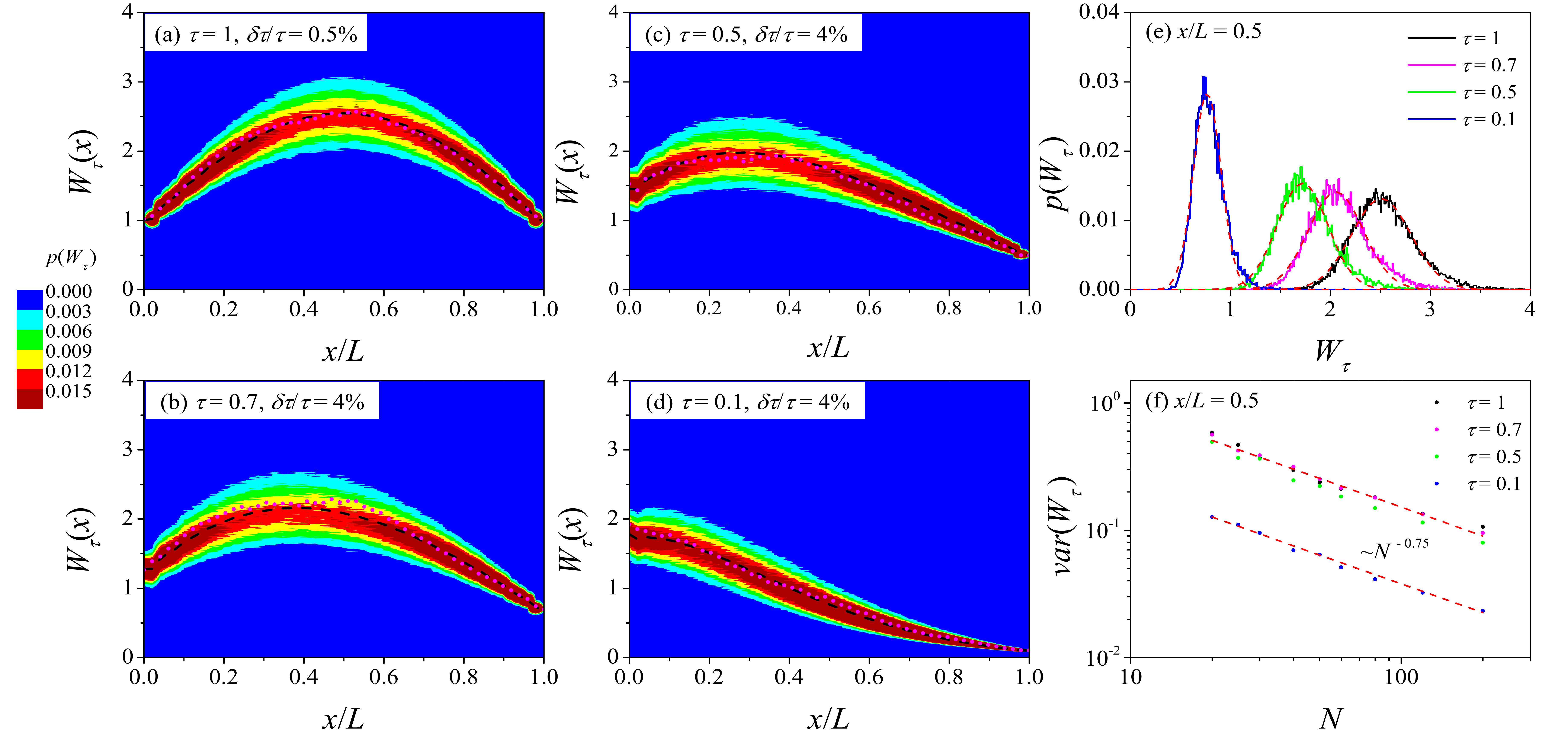}
\caption{(a-d) We simulate $W_\tau(x)$ for a large number of disorder configurations ($N=200$), and find a narrow distribution $p(W_\tau(x))$. The dashed line is the disorder average of $W_\tau(x)$; the profiles of a single disorder configuration at any large $N$ (e.g., $800$ for dotted line) converge to it. (e) $p(W_\tau(x))$ (histogram) is Gaussian (dashed line). (f) Its variance is $\sim N^{-0.75}$. ($L=50$)}
\label{fig:3}
\end{figure*}

To study the spatial structures of scattering eigenstates, we use the eigenchannel theory of waves \cite{Choi11,Tian15,Rotter17}.
We introduce the concept of eigenchannel in the following three steps. (i) Define the transmission matrix $t\equiv \{t_{ab}\}$ as $j_t=tj_i$, where $j_{i}$ and $j_{t}$ are the complex incoming and transmitted current amplitudes, respectively, which are vectors in the space spanned by the ideal waveguide modes $\varphi_{a}(y)$, with $a$ the mode index. The matrix element $t_{ab}=\sqrt{\tilde v_a\tilde v_b} \int\!\!\!\!\int dydy' \varphi_a(y)\varphi_b^*(y')G(x=L,y,x'=0,y')$. Here $G$ is Green's function of wave propagation, and $\tilde v_a$ is the group velocity of mode $a$. Then, we perform the singular value decomposition, $t=\sum_{n=1}^N u_n \sqrt{\tau_n} v_n^\dagger$, where the eigenvalue $\tau_n\in [0,1]$ decreases with $n$, and
$u_n
$ and $v_n
$ are mutual orthogonal unit vectors.
(ii) Make the extension: $t\rightarrow t(x)\equiv \{t_{ab}(x)\}$, where $t_{ab}(x)$ is obtained by replacing $G(x=L,y,x'=0,y')$ in $t_{ab}$ by $G(x,y,x'=0,y')$. This means that each $v_n$ gives an input field (in the transverse direction), that excites a
vector field, $E_{n}(x)
=t(x) v_n$, inside the medium, and $u_n$ is the corresponding output field, i.e., $E_{n}(L)=u_n$. (iii) The triple: $(v_n, E_n(x),\tau_n)$ defines an eigenchannel, with $n$ the channel index. Note that it depends on the disorder configuration and $\Omega$, but not on $c$. Each channel has a specific spatial profile (cf.~Fig.~\ref{fig:4}) of energy distribution, $|\sum_{a=1}^N E_{na}(x)\varphi^*_a(y)|^2$. Integrating over $y$ gives
\begin{equation}\label{eq:16}
    W_{\tau_n}(x)\equiv E_{n}^\dagger(x)\cdot E_{n}(x),
\end{equation}
which is called the eigenchannel structure. The channel's capability of transmitting waves is measured by $\tau_n$.

In the eigenchannel ($v_n$) basis, $j_i=\sum_{n=1}^N c_n v_n$.
From this we can find $E(x,y)=\sum_{a=1}^N(t(x)j_i)_a\varphi^*_a(y)$ and
the ($y$-integrated) energy density,
\begin{eqnarray}\label{eq:10}
    I(x;c)\!\equiv\! \int dy |E(x,y)|^2\!=\!\sum_{n,n'=1}^N \!c_n^*c_{n'} E_{n}^\dagger(x)\cdot E_{n'}(x),
\end{eqnarray}
for $0\leqslant x\leqslant L$. Equation (\ref{eq:10}) defines a family of functions over $S^{2N-1}$, which are labeled by $x$. Note that inside the medium the vectors $E_{n}(x)$ are not orthogonal.

To study Eq.~(\ref{eq:10}), we use the tool of the concentration of measure \cite{Milman87}. First of all, we introduce a mathematical object:
a continuous function $f: S^{2N-1}\rightarrow \mathds{R}$ is said to be $\sigma$-Lipschitz,
if $|f(c)-f(c')|\leqslant \sigma\|c-c'\|$ holds always. Here $\sigma$ is called the Lipschitz constant,
and $\|\cdot\|$ is some metric.
Then, we have\\
\\
\noindent L$\acute{\rm e}$vy's lemma. {\it  Let a continuous function $f: S^{2N-1}\rightarrow \mathds{R}$ be $\sigma$-Lipschitz with respect to the Euclidean metric. Then for any positive $\varepsilon$, the normalized area of the set in $S^{2N-1}$, over which the deviation $|f(c)-\int f(c)d\mu|$ exceeds $\varepsilon$ ($d\mu$ the normalized area element),
is $\leqslant 2e^{-\delta\varepsilon^2 N/\sigma^2}$.}\\
\\
\noindent In essence, it says that $f$ concentrates around its mean $\int fd\mu$. The degree of concentration is governed by the concentration function $2e^{-\frac{\delta \varepsilon^2 N}{\sigma^2}}$, the larger the ratio $N/\sigma^2$, the stronger the concentration effect.

In S1 of \cite{SM}, we show that $I(x;c)$ is Lipshitz continuous. More precisely, for $N\gg 1$,
\begin{equation}\label{eq:3}
    |I(x;c)-I(x;c')|\leqslant \sigma(x) \|c-c'\|,\quad \sigma(x)={\cal O}(1).
\end{equation}
Hereafter $\|\cdot\|$ refers to the Euclidean metric. This continuity property, especially that $\sigma(x)$ saturates at large $N$, relies heavily on the projection of incoming wave onto eigenchannel bases. Thanks to (\ref{eq:3}), we can apply the lemma to Eq.~(\ref{eq:10}). We find that the concentration function decays exponentially in $N$, i.e., the degree of concentration increases very fast with $N$. Thus for large $N$ and for almost every
$c$,
\begin{eqnarray}
I(x;c)=\int I(x;c)d\mu=\frac{1}{N}\sum_{n=1}^N W_{\tau_n}(x)\equiv I(x).
\label{eq:12}
\end{eqnarray}
Note that this result does not follow from the central limit theorem, but from the nice dependence of $I(x;c)$ on a large number of variables: $(c_1,c_2,\cdots,c_N)$. It has important implications. (i) An overwhelming number of incoming waves behave the same macroscopically: they pump energies into different channels with equal weight, giving rise to a steady state $I(x)$ universal with respect to $c$. Later we will see that $I(x)$ is universal with respect to disorder configurations also, and is a diffusive steady state. (ii) All the coherent terms ($n\neq n'$) in Eq.~(\ref{eq:10}) disappear. This implies that waves undergoing complex -- encoded in $\{E_n(x)\}$ -- scattering processes can simulate some effects of decoherence. (iii) $\{W_{\tau_n}(x)\}$ and $\{\tau_n\}$ completely determine $I(x)$.

L$\acute{\rm e}$vy's lemma also shows that there is an exceptional set of incoming waves, each of which excites a profile deviating significantly from $I(x)$.
However, the lemma is mute on the constructions of these waves. To explore the latter, the detailed knowledge on wave propagation is needed. In S3 of \cite{SM}, we show that the pattern of $\{|c_n|^2\}$ may suffice to identify these waves.

We proceed to study another important observable, the transmission $\tau(c)=j_t^\dagger \cdot j_t$. By using $j_i=\sum_{n=1}^N c_n v_n$, we find $\tau(c)=\sum_{n=1}^N |c_n|^2\tau_n$: this function can be shown to be $2$-Lipschitz (S4 of \cite{SM}). Then, from L$\acute{\rm e}$vy's lemma it follows that for large $N$ and for almost every $c$,
\begin{eqnarray}
\tau(c)=\int \tau(c)d\mu=\frac{1}{N}\sum_{n=1}^N \tau_n.
\label{eq:13}
\end{eqnarray}
Thus the transmission is universal with respect to $c$. As we will show later, for large $N$ it is universal with respect to disorder configurations also. Like Eq.~(\ref{eq:12}), Eq.~(\ref{eq:13}) does not follow from the central limit theorem.

We put the mathematical results into numerical test (see S5 of \cite{SM} for details). We simulate the wave propagation by using Eq.~(\ref{eq:15}) and obtain $I(x;c)$ for a large number of
$c$, drawn from a uniform distribution over $S^{2N-1}$. As shown in Fig.~\ref{fig:1} (a), $I(x;c)$ concentrates around $I(x)$ given by Eq.~(\ref{eq:12}),
and the deviation is very small; moreover, $I(x)$ falls linearly, in accordance with Fick's (Fourier's) law of particle (heat) diffusion \cite{Gaspard96,Lepri03}. As exemplified in (b,c),
at given $x$ the distribution of $I(x;c)$ is Gaussian, with a variance $\sim N^{-1}$. More precisely,
this variance is $\frac{(\sigma(x))^2}{N}$ up to an overall numerical factor,
consistent with the exponent of concentration function (S2 of \cite{SM}). These results
confirm that for large $N$, an overwhelming number of incoming waves behave the same macroscopically.
In particular, they excite a steady state $I(x)$ universal with respect to $c$.
Thanks to $I(x=L;c)=\tau(c)$, (c) confirms
Eq.~(\ref{eq:13}) as well. By simulations we also find exceptional incoming waves, the profiles of $I(x;c)$ excited by which deviate significantly from the linear fall (S3 of \cite{SM}).

Simulations further show that $I(x)$ is robust against the change in either disorder configuration
or $N (\gg 1)$. To study the underlying mechanism, we write Eq.~(\ref{eq:12}) as
\begin{eqnarray}
I(x)=\int_0^1 W_{\tau}(x) d\mu_N, \quad d\mu_N\equiv \frac{1}{N}\sum_{n=1}^N\delta(\tau-\tau_n)d\tau,
\label{eq:31}
\end{eqnarray}
where $\mu_N$ is the spectral distribution of $N\times N$ transmission matrix $t$ for a single disorder configuration. According to this, $I(x)$ is fully determined by $W_{\tau}(x)$ and $\mu_N$. So, to understand the robustness of $I(x)$, we explore the universality of $W_{\tau}(x)$ and $\mu_N$ in a single disorder configuration. (The universality of the average of $W_{\tau}(x)$ and $\mu_N$ over disorder configurations has been studied in literatures \cite{Tian15,Dorokhov84,Mello88}.) We begin with simulating $W_{\tau}(x)$ for a large number of disorder configurations. As shown in Fig.~\ref{fig:3} (a-d),
the disorder average of the profile $W_\tau(x)$, with eigenvalues in a narrow window of width $\delta\tau$ around $\tau$, is smooth (dashed line). Most importantly, we find that at given $x$ and $\tau$, fluctuations of $W_\tau(x)$ due to changes in disorder configurations are Gaussian (e), and as $N$ increases the average profile is not affected, but the variance decreases as $N^{-0.75}$ (f). This finding has an important consequence. That is, $W_\tau(x)$ converges at large $N$ to a profile, which is universal with respect to disorder configurations; thus, as confirmed by simulations (a-d), the disorder average of $W_\tau(x)$ at some large $N$ (dashed line) gives $W_\tau(x)$ of a single disorder configuration
at arbitrary large $N$ (dotted line).

\begin{figure}
\includegraphics[width=8.0cm] {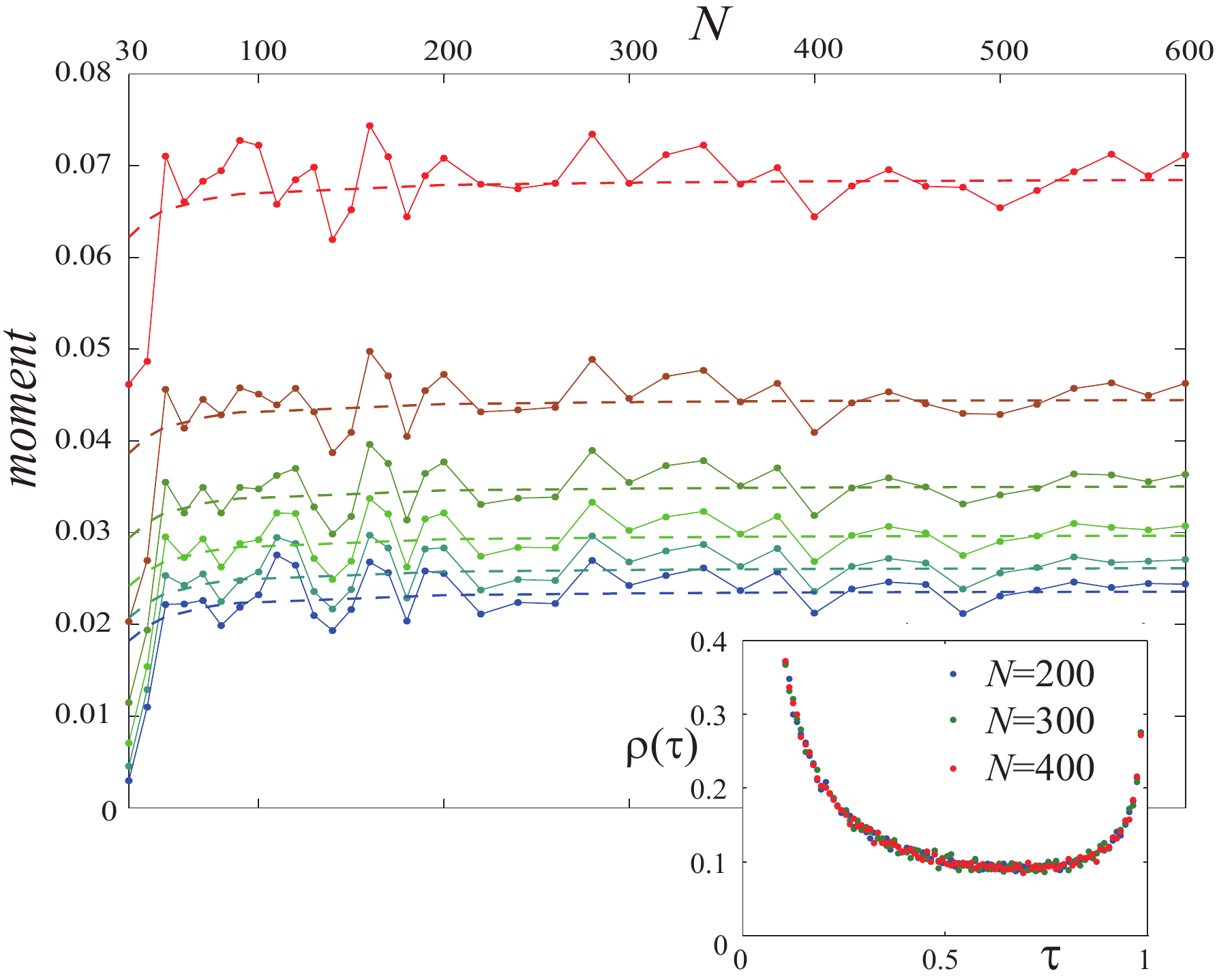}
\caption{Simulations show that the moments of $\mu_N$ (solid lines) converge to those of disorder averaged $\mu_{N}$
(dashed lines). From top to bottom the order of moment $p=1,2,\cdots,6$. Inset: at $N\gtrsim 200$ the average of $\mu_{N}$ over an ensemble of disorder configurations, generating $9\times 10^5$ eigenvalues totally, is stable, giving a limiting spectral density $\rho(\tau)$.}
\label{fig:2}
\end{figure}

To study the universality of $\mu_N$ in a single disorder configuration, we increase $N$ and obtain a sequence of $\mu_N$.
Simulations show that this sequence converge in moments (Fig.~\ref{fig:2}). Namely, the moments of $\mu_N$ converge to those of disorder averaged $\mu_N$ (main panel), and the latter has a limiting spectral distribution, with a density $\rho(\tau)$ (inset), i.e., $\lim_{N\rightarrow\infty}\int_0^1\tau^p d\mu_N=\int_0^1\tau^p \rho(\tau) d\tau\, (p=1,2,\cdots)$. Note that its left-hand side does not involve any disorder averaging. Since the support of spectral density is compact, the spectral distribution is uniquely determined by moments \cite{Tamarkin43}.
Then, by a theorem in probability theory \cite{Breiman92}, the convergence of $\mu_N$ in moments results in:
\begin{eqnarray}
I(x)\stackrel{N\gg 1}{=}\int_0^1 W_{\tau}(x) \rho(\tau)d\tau,
\label{eq:14}
\end{eqnarray}
where Eq.~(\ref{eq:31}) and the $\tau$-continuity of $W_{\tau}(x)$ were used. Note that by definition [cf.~Eq.~(\ref{eq:12})] $I(x)$ does not involve any disorder averaging. Equation (\ref{eq:14}) shows that $I(x)$ converges to the universal profile: $\int_0^1 W_{\tau}(x) \rho(\tau)d\tau$ at large $N$, and thus explains the robustness of $I(x)$. Setting $x=L$ in Eq.~(\ref{eq:14}), we find that $\tau(c)=\int_0^1 \tau\rho(\tau)d\tau$ for large $N$, and thus is universal not only with respect to $c$, but also to disorder configurations.

Our findings provide a new angle for understanding some fundamental issues of nonequilibrium statistical mechanics, notably the origin of decoherence and irreversibility. They may also have practical applications, such as controlling propagation of light waves in disordered media \cite{Rotter17}, where one often deals with a single disorder configuration. Finally, we leave the important issues of nonequilibrium temporal structures and interaction effects for future studies.

\begin{acknowledgements}
C.T. is supported by the National Natural Science Foundation of China (No. 11535011 and No. 11647601).\\
\noindent $^*$ Electronic address: ct@itp.ac.cn\\
\noindent $^\dagger$ These two authors contribute equally to this work.
\end{acknowledgements}

\clearpage

\renewcommand{\thesection}{S\arabic{section}}
\renewcommand{\thesubsection}{\thesection.\arabic{subsection}}
\renewcommand{\theequation}{S\arabic{equation}}
\renewcommand{\thefigure}{S\arabic{figure}}

\setcounter{page}{1}
\setcounter{equation}{0}
\setcounter{figure}{0}

\begin{widetext}

\begin{center}
{\bf Supplemental Materials}
\end{center}

\noindent {\bf S1. The Lipschitz constant of $I(x;c)$}

It is easy to see that the function $I(x;c): S^{2N-1}\rightarrow \mathds{R}$ is differentiable. In this section, we further show that it is Lipschitz and for $N\gg 1$ the Lipschitz constant is essentially independent of $N$.\\
\\
\noindent {\bf S1.1. The Lipschitz constants with respect to the Euclidean and geodesic metrics}

Note that $S^{2N-1}$ is equipped with a ``natural'' metric, namely, the geodesic metric. Let $f: S^{2N-1}\rightarrow \mathds{R}$ be a Lipschitz function with respect to this metric. Its Lipschitz constant (with respect to this metric) can be chosen as the least upper bound of the set, $\{\|\nabla_c f\|:c\in S^{2N-1}\}$, and is denoted as ${\rm sup}\|\nabla_c f\|$ (cf.~the remark below). Here $\nabla_c$ is the tangential derivative. Recall that $\|\cdot\|$ is the Euclidean metric. What is the Lipschitz constant with respect to the Euclidean metric? The answer is given by the following lemma.\\

{\it Let a continuous function $f: S^{2N-1}\rightarrow \mathds{R}$ be $\sigma$-Lipschitz (with respect to the Euclidean metric). Then $\sigma$ can be chosen as $\frac{\pi}{2}{\rm sup}\|\nabla_c f\|$.}\\

{\it Proof.} Suppose that $\gamma (c,c')$ is the shorter geodesic connecting $c,c'\in S^{2N-1}$.  Then
\begin{equation}\label{eq:17}
    |f(c)-f(c')|=\left|\int_{\gamma (c,c')}ds \frac{\partial f}{\partial s}\right|\leqslant \int_{\gamma (c,c')}ds \left|\frac{\partial f}{\partial s}\right|,
\end{equation}
where $ds$ is the line element of $\gamma (c,c')$. Because of $|\partial f/\partial s|\leqslant \|\nabla_c f\|$,
we have
\begin{equation}\label{eq:19}
    \int_{\gamma (c,c')}ds \left|\frac{\partial f}{\partial s}\right|\leqslant \int_{\gamma (c,c')}ds \|\nabla_c f\|\leqslant {\rm sup}\|\nabla_c f\| \int_{\gamma (c,c')}ds.
\end{equation}
Note that $\int_{\gamma (c,c')}ds$ is the geodesic distance of $\gamma$. It satisfies
\begin{equation}\label{eq:27}
\int_{\gamma (c,c')}ds\leqslant \frac{\pi}{2}\|c-c'\|.
\end{equation}
Combining the inequalities (\ref{eq:17}), (\ref{eq:19}) and (\ref{eq:27}), we have
\begin{equation}\label{eq:28}
    |f(c)-f(c')|\leqslant \frac{\pi}{2}{\rm sup}\|\nabla_c f\|\|c-c'\|.
\end{equation}
This completes the proof of theorem. $\Box$

{\it Remark.} The inequalities (\ref{eq:17}) and (\ref{eq:19}) show that the Lipschitz constant with respect to the geodesic metric can indeed be chosen as ${\rm sup}\|\nabla_c f\|$.\\
\\
\begin{figure}
\includegraphics[width=12.7cm] {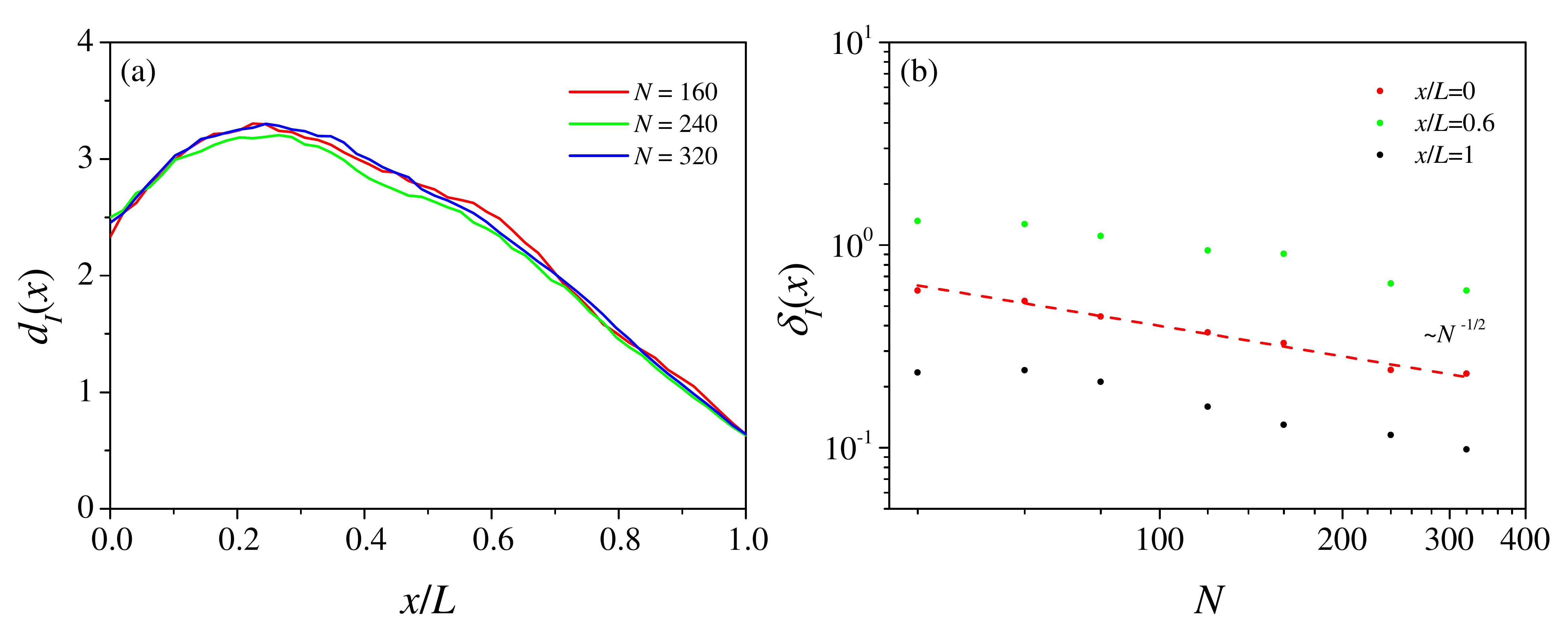}
\caption{Numerical simulations show that the profile $d_I(x)$, obtaining by averaging $\|\nabla_c I(x;c)\|$ with respect to $c$, converges at large $N$ (a), and at different $x$ the maximal deviation $\delta_I(x)$ from $d_I(x)$ decreases with $N$ as $N^{-1/2}$ (b).}
\label{fig:S1}
\end{figure}

\noindent {\bf S1.2. The method of numerically calculating $\|\nabla_c I\|$}

To calculate the Lipschitz constant of $I(x;c)$ analytically is a difficult task, since it requires the detailed knowledge of the field profiles of eigenchannels. Therefore, we resort to numerical analysis. First of all, we need to numerically calculate $\|\nabla_c I\|$. To this end, we note that by using Eq.~(\ref{eq:10}), it can be shown straightforwardly
\begin{equation}\label{eq:29}
    \|\nabla_c I\|=2\sqrt{\sum_{n=1}^N\frac{\partial I}{\partial c_n}\frac{\partial I}{\partial c_n^*}-I^2},
\end{equation}
where
\begin{equation}\label{eq:30}
    \frac{\partial I}{\partial c_n}=\sum_{n'=1}^N \!c_{n'}^* E_{n'}^\dagger(x)\cdot E_{n}(x),\quad \frac{\partial I}{\partial c_n^*}=\sum_{n'=1}^N \!c_{n'} E_{n}^\dagger(x)\cdot E_{n'}(x).
\end{equation}
This defines a new function of $c$. It can be numerically calculated rather easily, because thanks to Eq.~(\ref{eq:30}) it does not involve any derivatives. Indeed, for given incoming wave, $c$ we can simulate Eq.~(\ref{eq:15}) and obtain the set of $\{E_n(x)\}$. Then we substitute the result as well as the value of $c$ into Eqs.~(\ref{eq:29}) and (\ref{eq:30}). This gives the numerical value of $\|\nabla_c I\|$ at given $c$. The great advantage of this method is that we do not need to use the numerical differentiation to calculate $\|\nabla_c I\|$.\\
\\
\noindent {\bf S1.3. Numerical result}

We draw $c$ from a uniform distribution over $S^{2N-1}$. The algorithm of generating this random $c$ is described in Sec.~S5. Then we numerically calculate the derivative $\nabla_c I(x;c)$ by using the method described in Sec.~S1.2.
This gives an $x$-profile of $\|\nabla_c I(x;c)\|$.
We repeat this procedure for $M=10000$ times and perform the average of the profiles of $\|\nabla_c I(x;c)\|$. The average profile $d_I(x)=\frac{1}{M}\sum_{k=1}^M\|\nabla_c I(x;c)\||_{c=c_k}$. In Fig.~\ref{fig:S1} (a) we present the result of $d_I(x)$ for different $N$. We see that the profile of $d_I(x)$ converges at large $N$. Furthermore,
for different but fixed $x$, we calculate the maximal deviation $\delta_I(x)$ from $d_I(x)$. As shown in Fig.~\ref{fig:S1} (b), $\delta_I(x)$ scales with $N$ as $N^{-1/2}$ at every $x$. So, $\delta_I(x)$ decreases with $N$. This implies that $\|\nabla_c I(x;c)\|=d_I(x)$ for large $N$,
and is independent of $c$. Since the Lipschitz constant with respect to the geodesic metric of $I(x;c)$ is ${\rm sup} \|\nabla_c I(x;c)\|$, by using the lemma proved in Sec.~S1.1 we find
\begin{equation}\label{eq:S1}
    \sigma(x)=\frac{\pi}{2} d_I(x),\quad {\rm for}\, N\gg 1.
\end{equation}
This gives the result (\ref{eq:3}).
\\
\\
\begin{figure*}[h]
\includegraphics[width=17.5cm] {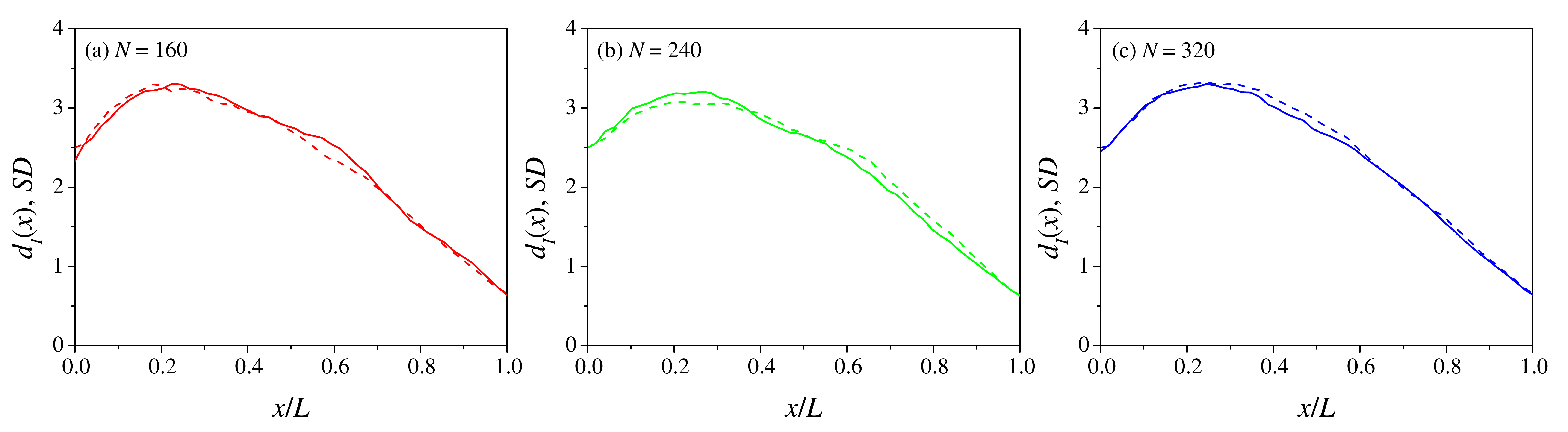}
\caption{Simulations show that for large $N$ the profile of rescaled SD (dashed line) collapses into that of $d_I(x)$ (solid line).}
\label{fig:S3}
\end{figure*}

\noindent {\bf S2. The variance of
$\boldsymbol{I(x;c)}$}

As already shown in Fig.~\ref{fig:1}(b-c), simulations confirm that at every $x$, the fluctuations of $I(x;c)$ due to the change in the incoming wave $c$ are Gaussian, and the variance scales with $N$ as $N^{-1}$.
Here we present the complete simulation results of this variance. We first rescale the simulated standard deviation (SD) at every $x$ by its value at $x=0$. Then, the rescaled SD is multiplied by a numerical factor of $2.5$. The factor is close to $d_I(0)$ (cf.~Fig.~\ref{fig:S1}).
As shown in Fig.~\ref{fig:S3}, the ensuing SD profile and the profile of $d_I(x)$ are identical.
This implies that the variance of $I(x;c)$ is $\frac{(d_I(x))^2}{N}
$, up to an overall numerical factor. Because of Eq.~(\ref{eq:S1}), this variance is $\frac{(\sigma(x))^2}{N}$ essentially.

Furthermore, we show below that the simulation results of $p(I(x;c))$ are also consistent with the concentration function from L$\acute{\rm e}$vy's lemma. Since $p(I(x;c))$ is Gaussian, with the variance given above, we find that at given $x$, the probability for the deviation of $I(x;c)$ from its $c$-averaging $I(x)$ to exceed $\varepsilon$ is
\begin{equation}
    {\rm Pr.} \{|I(x;c)-I(x)|>\varepsilon\}=\int_{|I(x;c)-I(x)|>\varepsilon}p(I(x;c))dI(x;c)={\rm ercf}\left(-\frac{\delta' \varepsilon^2 N}{(d_I(x))^2}\right),
\label{eq:S4}
\end{equation}
where ${\rm ercf}(t)\equiv \frac{2}{\sqrt{\pi}}\int_t^\infty e^{-s^2}ds$ is the complementary error function, and $\delta'>0$ is an absolute constant. On the other hand, this probability is the normalized area of the set in $S^{2N-1}$, over which $|I(x;c)-I(x)|>\varepsilon$. Applying the inequality: ${\rm ercf}(t)\leqslant 2e^{-t^2}$ to Eq.~(\ref{eq:S4}), we find that this area is
$\leqslant 2e^{-\frac{\delta' \varepsilon^2 N}{(d_I(x))^2}}$.
With the substitution of Eq.~(\ref{eq:S1}), we reproduce the concentration function from L$\acute{\rm e}$vy's lemma.\\
\\
\begin{figure*}[h]
\includegraphics[width=17.5cm] {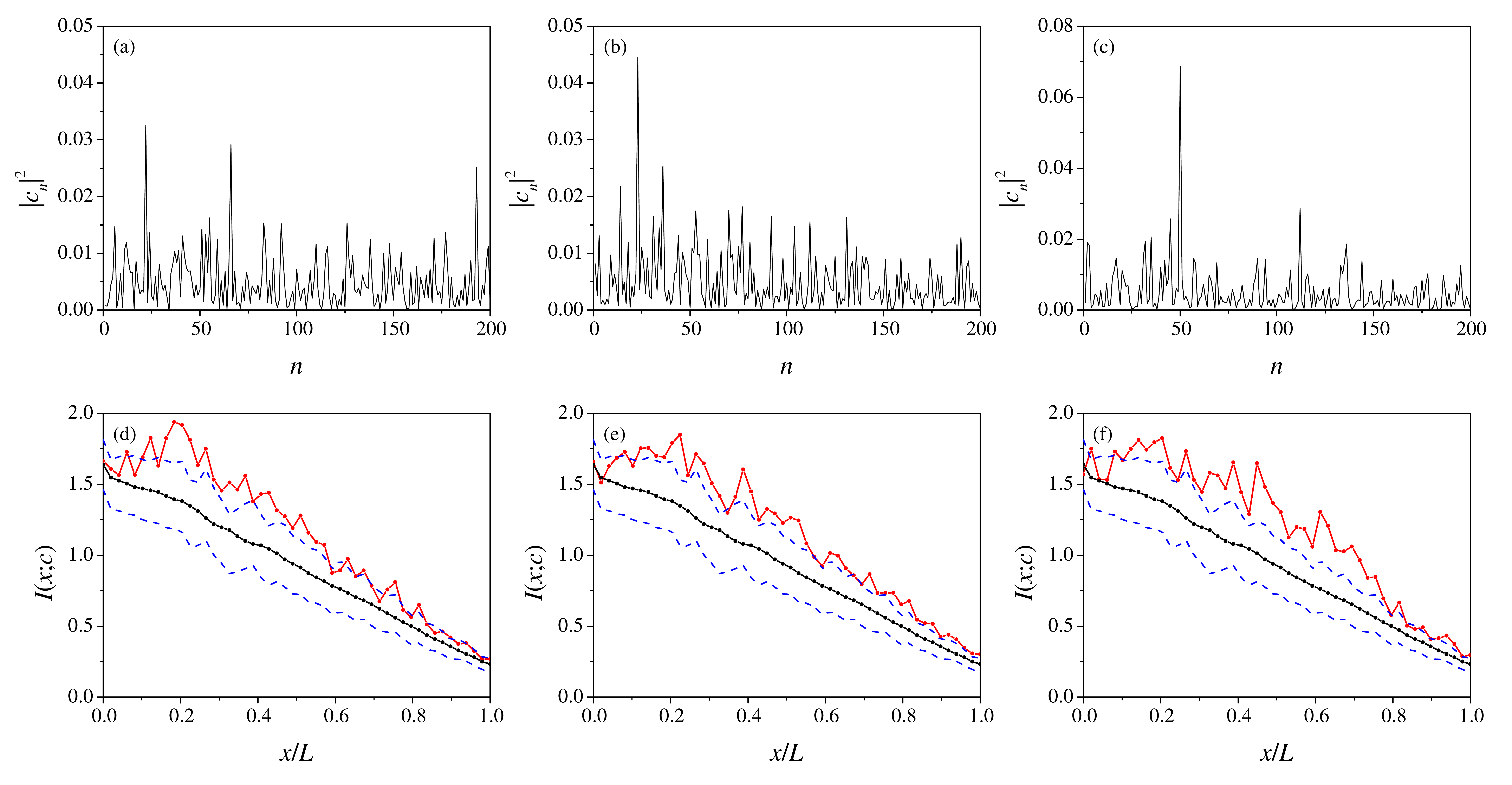}
\caption{Examples of the eigenchannel components of exceptional incoming waves (a-c). Corresponding energy density profiles are represented by red solid lines in (d-f). In (d-f) the blue dashed lines are reference profiles, which deviate at every $x$ from $I(x)$ (black solid line) by twice SD.}
\label{fig:S2}
\end{figure*}

\noindent {\bf S3. Exceptional incoming waves}

L$\acute{\rm e}$vy's lemma, in combination with Eq.~(\ref{eq:S1}), shows that there is an exceptional set of incoming waves, whose normalized area is exponentially small in $N$. Each incoming wave $c$ in this set excites a profile $I(x;c)$, which deviates significantly from $I(x)$. However, L$\acute{\rm e}$vy's lemma is mute on the structure of these incoming waves. Representing the incoming wave in the eigenchannel ($v_n$) basis provides the possibility of unveiling this structure. Indeed, as shown in Fig.~\ref{fig:3} (a-d) already, this set includes the incoming wave $j_i=v_n$, since in this case the excited $I(x;c)$ is just the eigenchannel structure $W_{\tau_n}(x)$, whose profile is completely different from a linear fall in general. In order to find (rather than construct theoretically) more general examples of exceptional incoming waves, we repeat the numerical experiments as described in the caption of Fig.~\ref{fig:1}. We set $N=200$ so that we can increase the number of incoming waves by an order. Then, we pick up the incoming waves giving rise to a profile $I(x;c)$ that significantly deviates from $I(x)$, and examine the components of the incoming wave. We find that for such waves there are always few components of $|c_n|^2$, with $n$ corresponding to high and (or) low transmission eigenchannels, whose values are relatively high. Some examples are shown in Fig.~\ref{fig:S2}.\\
\\

\noindent {\bf S4. Lipschitz continuity of $\tau(c)$}

In this section, we need to use the Cauchy-Schwartz inequality,
\begin{equation}\label{eq:22}
    \left(\sum_{i=1}^m a_ib_i\right)^2\leqslant \left(\sum_{i=1}^m a_i^2\right)\left(\sum_{i=1}^m b_i^2\right), \quad \forall\, a_i,b_i\in \mathds{R},
\end{equation}
and the triangle inequality,
\begin{equation}\label{eq:25}
    \|a+b\|\leqslant \|a\|+\|b\|, \quad \forall\, a,b\in \mathbb{C}^n.
\end{equation}
For the function $\tau:\, c\in S^{2N-1}\mapsto \tau(c)\in \mathds{R}$, we have the following theorem:\\

{\it $\tau(c)$ is $2$-Lipschitz, i.e.,
\begin{eqnarray}
|\tau(c)-\tau(c')|\leqslant 2\|c-c'\|.
\nonumber
\end{eqnarray}
}

{\it Proof.} By using $\tau(c)=\sum_{n=1}^N |c_n|^2\tau_n$, we obtain
\begin{eqnarray}
|\tau(c)-\tau(c')|=\left|\sum_{n=1}^N \left(|c_n|^2-|c'_n|^2\right)\tau_n\right|.
\label{eq:20}
\end{eqnarray}
Substituting $c_n\equiv a_n+ib_n,c'_n\equiv a'_n+ib'_n$, where $a_n,b_n,a'_n,b'_n\in \mathds{R}$, into it, we obtain
\begin{eqnarray}
|\tau(c)-\tau(c')|=\left|\sum_{n=1}^N \left(\left(a_n-a'_n\right)\left(a_n+a'_n\right)+\left(b_n-b'_n\right)\left(b_n+b'_n\right)\right)\tau_n\right|.
\label{eq:21}
\end{eqnarray}
Using the Cauchy-Schwartz inequality (\ref{eq:22}), we obtain
\begin{eqnarray}
|\tau(c)-\tau(c')|&\leqslant&\left(\sum_{n=1}^N \left(a_n-a'_n\right)^2+\sum_{n=1}^N \left(b_n-b'_n\right)^2\right)^{\frac{1}{2}}\left(\sum_{n=1}^N \left(a_n+a'_n\right)^2\tau_n^2+\sum_{n=1}^N \left(b_n+b'_n\right)^2\tau_n^2\right)^{\frac{1}{2}}\nonumber\\
&=&\left(\sum_{n=1}^N |c_n-c_n'|^2\right)^{\frac{1}{2}}\left(\sum_{n=1}^N |c_n+c_n'|^2\tau_n^2\right)^{\frac{1}{2}}\nonumber\\
&=&\|c-c'\|\left(\sum_{n=1}^N |c_n+c_n'|^2\tau_n^2\right)^{\frac{1}{2}}.
\label{eq:23}
\end{eqnarray}
The remaining task is to bound the second factor in the last line.

Because of $\tau_n\leqslant 1$ we have
\begin{eqnarray}
\left(\sum_{n=1}^N |c_n+c_n'|^2\tau_n^2\right)^{\frac{1}{2}}\leqslant \left(\sum_{n=1}^N |c_n+c_n'|^2\right)^{\frac{1}{2}}=\|c+c'\|.
\label{eq:24}
\end{eqnarray}
Applying the triangle inequality (\ref{eq:25}) to it we obtain
\begin{eqnarray}
\left(\sum_{n=1}^N |c_n+c_n'|^2\tau_n^2\right)^{\frac{1}{2}}\leqslant 2.
\label{eq:26}
\end{eqnarray}
This completes the proof of the theorem. $\Box$\\
\\

\noindent {\bf S5. The method of numerical simulations}

We simulate the wave propagation by using Eq.~(\ref{eq:15}), where $\delta \epsilon (x,y)$ is drawn from a uniform distribution over an interval $[-\delta \epsilon_0,\delta \epsilon_0]$. Here $\delta \epsilon_0\in (0,1)$ governs the disorder strength and is set to $0.97$ in simulations. The values of $\delta \epsilon (x,y)$ at distinct spatial points are chosen in the same way and independently. For simulations, Eq.~(\ref{eq:15}) is discretized on a square grid, with the grid spacing being the inverse wave number in the ideal waveguide, and the length of medium is defined as the number of discrete points in $x$-direction. We solve this equation by using the recursive Green's function method [A. MacKinnon, {\it Z. Phys. B} \textbf{59}, 385 (1985)]. We first calculate $G(x=L,y,x'=0,y')$, from which we obtain the transmission matrix $t$. By performing the singular value decomposition of $t$, we obtain the input field $v_n$ and the eigenvalue $\tau_n$ of the $n$th eigenchannel. Next, we calculate $G(x,y,x'=0,y')$, from which we obtain the matrix $t(x)$. Then the wave field in the interior of the medium is given by $E(x,y)=\sum_{a=1}^N (t(x)j_i)_a\varphi^*_a(y)$. For a single disorder configuration, small oscillations in $x$ are often superposed on non-oscillating backgrounds. These oscillations occur in the wavelength scale and are unimportant. They are removed by performing the local (in $x$) average over a window with a width of wavelength.

To draw a point $c\equiv(c_1,c_2,\cdots,c_N)$ from a uniform distribution over $ S^{2N-1}$, we use the standard method [G. Marsaglia, {\it Ann. Math. Stat.} \textbf{43}, 645 (1972)]. First, we generate $2N$ independent standard normal random variables, $(a_n,b_n)$ with $n=1,2,\cdots, N$. Secondly, we normalize each $a_n$ ($b_n$) by $\sqrt{\sum_{n=1}^N(a_n^2+b_n^2)}$. Define the normalized $a_n$ ($b_n$) as the real (imaginary) part of $c_n$. We generate a desired random point $c$.

\clearpage
\end{widetext}

\end{document}